\documentclass{aa}

\begin{document}

\title{New measurements of magnetic fields of roAp stars with FORS\,1 at the VLT
\thanks{Based on observations obtained at the European Southern Observatory, Paranal, Chile (ESO programme No.~269.D-5044)}}

\author{S. Hubrig\inst{1}\and T. Szeifert \inst{1}\and M. Sch\"oller
\inst{1}\and G. Mathys \inst{1} \and D.W. Kurtz \inst{2}}

\institute{European Southern Observatory, Casilla 19001, Santiago 19, Chile
\and Centre for Astrophysics, University of Central Lancashire, Preston, PR1 2HE, UK
}

\date{Received xx/Accepted yy}

\offprints{S. Hubrig}

\titlerunning{Magnetic fields of roAp stars}
\authorrunning{Hubrig et al.}

\abstract{Magnetic fields play a key role in the pulsations of rapidly oscillating Ap (roAp) 
stars since they are a necessary ingredient of all pulsation excitation mechanisms proposed so 
far. This implies that the proper understanding of the seismological behaviour of the roAp stars 
requires knowledge of their magnetic fields. However, the magnetic fields of the roAp stars are 
not well studied. Here we present new results of measurements of the mean longitudinal field of 
14 roAp stars obtained from low resolution spectropolarimetry with FORS1 at the VLT. 


}

\maketitle

\section{Introduction}\label{sec1}
Helioseismology has proved a powerful tool to investigate important basic properties of the 
solar interior (Gough \cite{G00}). Likewise, asteroseismology has the potential to provide new 
insights into the physics of stellar interiors. Among the most promising objects that can be 
studied through this technique are the roAp stars. These are cool chemically peculiar stars 
which pulsate in high-overtone ($n\gg{}l$), low-degree ($l\le3$) $p$-modes with periods from 
about 6 to 15 min and typical photometric $B$ amplitudes of a few mmag. There are 32 such stars 
currently known. Detailed reviews of the roAp stars have been published by Kurtz (\cite{K90}), 
Matthews (\cite{M91}), Martinez \& Kurtz (\cite{MK95}) and Kurtz (\cite{K03}).

The pulsations of the roAp stars are strongly affected by their magnetic fields (Bigot et al. 
\cite{B00}; Cunha \& Gough \cite{CG00}). Of particular importance is the recent work by Bigot 
\& Dziembowski (\cite{BD02}). Their theory of the interaction of rotation, pulsation and the 
magnetic field suggests a new model, the improved oblique pulsator model, that is a significant 
departure from the standard oblique pulsator model for roAp stars. They suggest that the light 
variations are caused by a pulsation mode in which the stellar surface moves in a plane that is 
inclined to both the rotation and the magnetic axes of the star. The displacement vector 
describes an ellipse in that plane with the pulsation period, and the whole pattern rotates 
with the rotation of the oblique magnetic field.

Obviously, magnetic fields play a key role in the pulsations, which can only be properly 
interpreted if the strength and structure of these fields are known. Despite of the importance 
of magnetic fields for the proper understanding of pulsational properties of roAp stars, these 
fields have scarcely been studied until now. Among 32 roAp stars 20 definitely or very probably 
have a magnetic field (Mathys \cite{M03}), but of these 20, only three stars, HD\,12098,
HD\,24712 and HD\,83368, have been studied throughout their rotation periods (L\"uftinger 
et al. \cite{LA03}; Mathys \& Hubrig \cite{MH97}). Here we present new results of measurements 
of the mean longitudinal field of 14 roAp stars obtained from low resolution spectropolarimetry 
with FORS1 at the VLT. 

\section{Observations and data reduction}\label{sec2}

The observations reported here have been carried out between July and September 2002 at the 
European Southern Observatory with FORS\,1 (FOcal Reducer low dispersion Spectrograph) mounted 
on the 8-m Melipal telescope of the VLT. This multi-mode instrument is equipped with 
polarization analyzing optics comprising super-achromatic half-wave and quarter-wave phase 
retarder plates, and a Wollaston prism with a beam divergence of 22$\arcsec$ in standard 
resolution mode. For 13 stars we used the GRISM\,600B in the wavelength range 
3480--5890\,\AA{} to cover all hydrogen Balmer lines from H$\beta$ to the Balmer jump 
(Appenzeller et al. \cite{AF98}). 
One star, HD\,6532, has been observed with GRISM\,600R in the region which 
includes H$\alpha$ and H$\beta$, from 4770 to 6900\,\AA. Both grisms have 
600 grooves mm$^{-1}$; with the narrowest available slit width of 0$\farcs$4 they give a 
spectral resolving power of R$\sim $2000 and R$\sim$2900, respectively. 

Wavelength calibrations were taken during day time for the two different retarder waveplate 
setups ($\alpha$=$+$45$^\circ$ and $-$45$^\circ$) that were used for the observations. 
Wavelength calibration was performed by associating with each individual science spectrum the 
calibration frame obtained with the similar orientation of the retarder waveplate. Ordinary and 
extraordinary beams were independently calibrated with the corresponding beams of the reference 
spectrum. As has been previously shown by Landstreet (\cite{L82}), in the weak field regime, 
the mean longitudinal field can be derived from the difference between the circular 
polarizations observed in the red and blue wings of the hydrogen line profiles using the formula 
\begin{equation}
\frac{V}{I} = -\frac{g_{\rm eff} e \lambda^2}{4\pi{}m_ec^2}\ \frac{1}{I}\ \frac{{\rm d}I}{{\rm d}\lambda} \left<{\cal B}_z\right>,
%\label{eq:one}
\end{equation}
where $V$ is the Stokes parameter which measures the circular polarization, $I$ is the intensity 
in the unpolarized spectrum, $g_{\rm eff}$ is the effective Land\'e factor, $e$ is the electron 
charge, $\lambda$ is the wavelength, 
% expressed in \AA{}, 
$m_e$ the electron mass, $c$ the speed of light, ${{\rm d}I/{\rm d}\lambda}$ is the derivative 
of Stokes $I$, 
and $\left<{\cal B}_z\right>$ is the mean longitudinal field.
% expressed in Gauss. 
% GM: with the purely symbolic form of Eq. (1) that you present here,
% it does not matter in which units the various quantities are expressed,
% provided that these units are mutually consistent (this would be 
% different if you had replaced (e/(4\,\pi\,\m_e\,c^2) by 
% the numerical value 4.67 E-13, as I often do).
To minimize the cross-talk effect we executed the sequence $+45-45$, $+45-45$, $+45-45$ etc. 
and calculated the values $V/I$ using: 
\begin{equation}
\frac{V}{I} = \frac{1}{2} \left\{ \left( \frac{f^{\rm o} - f^{\rm e}}{f^{\rm o} + f^{\rm e}} \right)
_{\alpha=-45^{\circ}} - \left( \frac{f^{\rm o} - f^{\rm e}}{f^{\rm o} + f^{\rm e}} \right)_{\alpha=+
45^{\circ}} \right\},
%\label{eqn:two} 
\end{equation}
where $\alpha$ gives the position angle of the retarder waveplate and $f^{\rm o}$ and 
$f^{\rm e}$ are ordinary and extraordinary beams, respectively. Stokes $I$ values have been 
obtained from the sum of the ordinary and extraordinary beams. In our calculations we assumed a 
Land\'e factor 
$g_{\rm eff}$=1 for hydrogen lines. More details of the observing technique are given by Bagnulo 
et al. (\cite{B02}) and Hubrig et al. (\cite{H03}).
%To derive $\left< {\cal B}_z \right>$ a least-squares technique has been used to minimize the 
% expression
%\begin{displaymath}
%\chi^2 = \sum_i \frac{(y_i - \left< {\cal B}_z \right> x_i - b)^2}{\sigma_i^2}
%\end{displaymath}
%where, for each spectral point $i$, $y_i = (V/I)_i$,
%$x_i = -(g_{\rm eff} e \lambda_i^2)/(4\pi{}m_ec^2)\ (1/I\ \times\ {\rm d}I/{\rm d}\lambda)_i$
%and $b$ is a constant term that, assuming that Eq.~(1) is correct, approximates the fraction of 
%instrume%ntal polarization not removed after the application of Eq.~(2) to the observations. For 
%each spectral po%int $i$, the derivative of Stokes $I$ with respect to the wavelength was 
%evaluated as 
%\begin{displaymath}
%\left( \frac{{\rm d}I}{{\rm d}\lambda} \right)_{\lambda=\lambda_i} = \frac{N_{i+1}-N_{i-1}}{\lamb%da_{i+1}-\lambda_{i-1}},
%\end{displaymath}
%where $N_i$ is the photon count at wavelength $\lambda_i$. In our calculations we assumed a 
%Land\'e factor g$_{\rm eff}$=1 for hydrogen lines and g$_{\rm eff}$=1.25 for metal lines. 
%Table~\ref{tab:balmer} lists the wavelength ranges
%corresponding to the hydrogen Balmer lines for which the Land\'e factor has been set to 1.

\begin{table}
\caption{
\label{tab:balmer}
Wavelength ranges around the hydrogen Balmer lines. 
}
\begin{center}
\begin{tabular}{cc}
Line(s) & Wavelength range [\AA{}] \\
\hline
H$_{16}$-H$_{10}$ & 3701 -- 3801 \\
H$_{9}$ & 3820.5 -- 3852.5 \\
H$_{8}$ & 3870.2 -- 3910.2 \\
H$\epsilon$ & 3941.2 -- 4001.2 \\
H$\delta$ & 4082.9 -- 4122.9 \\
H$\gamma$ & 4311.7 -- 4371.7 \\
H$\beta$ & 4812.7 -- 4912.7 \\
H$\alpha$ & 6512 -- 6612 \\
\end{tabular}
\end{center}
\end{table}
The errors of the measurements of the polarization have been determined from photon counting 
statistics and have been converted to errors of field measurements. For each roAp star we 
usually took 4 to 13 continuous series of 
2 exposures with the retarder waveplate oriented at different angles. The spectropolarimetric 
capability of the FORS\,1 
instrument in combination with the large light collecting power of the VLT allowed us to achieve 
a  
S/N ratio up to 2000 per pixel in the one--dimensional spectrum, as required to detect low 
polarization 
signatures produced in the spectral lines by longitudinal magnetic fields of the order of a few 
hundred Gauss. 

\section{Results}\label{sec3}

The basic data of our sample of roAp stars are presented in Table~\ref{tab:overview}. The 
columns are, 
in order, the HD number of the star, another identifier, the visual magnitude, the spectral 
type as it appears 
in the catalogue of Renson et al. (\cite{R91}), the pulsation period, the heliocentric Julian 
date of 
mid--observation, number of series N and the longitudinal 
magnetic field $\left<{\cal B}_z\right>$ and its estimated uncertainty $\sigma{}_z$. 
The star HD~\,9563 shows the highest uncertainty of the 
magnetic field measurements ($\sigma{}_z= 325$\,G) because the exposure was taken in 
deteriorating 
weather conditions. Among the roAp stars studied, the rotation period is known only for the star 
HD\,6532 
(P$_{\rm rot}=1.\!\!{^d}94497$; Kurtz et al. \cite{K96}).

\begin{table*}
\caption{
\label{tab:overview}
Basic data of studied roAp stars and their magnetic fields measured with the FORS\,1 instrument 
on the VLT UT3 (Melipal).}
\begin{center}
\begin{tabular}{rlrlrlcc}
HD & Other & V & Sp.\ Type & P$_{\rm puls}$ & 
HJD & N&$\left<{\cal B}_z\right>$ \\
 & identifier & & & [min]  & {\footnotesize 2452000+}& & [G] \\
\hline
6532 & CD -27$^{\circ}$355 & 8.4 & A3 Sr Cr & 6.9 
& 531.915 &13& 215 $\pm$ 158 \\
9289 & BD -11$^{\circ}$286 & 9.4 & A3 Sr Eu &10.5 
& 519.751 &4& 654 $\pm$ 130 \\
12932 & BD -19$^{\circ}$384 & 10.4 & A5 Sr Eu & 11.6 
& 517.884 &5& 638 $\pm$ 169 \\
99563 & BD -08$^{\circ}$3173 & 8.2 & F0 Sr & 11.2  
& 494.483 & 5&$-$688 $\pm$ 325 \\
122970 & BD +06$^{\circ}$2827 & 8.3 & F0 & 11.1 
& 494.500 &5& 216 $\pm$ 125 \\
161459 & CD -51$^{\circ}$11145 & 10.4 & A2 Eu Sr Cr & 12.0  
& 476.622 & 4&$-$1755 $\pm$ 224 \\
185256 & CD -30$^{\circ}$17252 & 9.9 & F0 Sr Eu & 10.2  
& 476.676 &4& $-$706 $\pm$ 180 \\
190290 & CPD -79$^{\circ}$1062 & 9.9 & A0 Eu Sr & 7.3 
& 494.561 & 5&3220 $\pm$ 164 \\
190290 & CPD -79$^{\circ}$1062 & 9.9 & A0 Eu Sr & 7.3 
& 498.529 & 5&3250 $\pm$ 248 \\
193756 & CPD -52$^{\circ}$11681 & 9.2 & A9 Sr Cr Eu & 13.0 
& 498.575 &10& $-$193 $\pm$ 136 \\
196470 & BD -18$^{\circ}$5731 & 9.8 & A2 Sr Eu & 10.8 
& 476.732 &4& 1474 $\pm$ 202 \\
203932 & CD -30$^{\circ}$18600 & 8.8 & A5 Sr Eu & 5.9  
& 498.615 &4& $-$267 $\pm$ 143 \\
213637 & BD -20$^{\circ}$6447 & 9.6 & F1 Eu Sr Cr & 11.5  
& 498.653 &8& 740 $\pm$ 141 \\
217522 & CD -45$^{\circ}$14901 & 7.5 & A5 Sr Eu Cr & 13.7 
& 498.687 &4& $-$725 $\pm$ 176 \\
218495 & CPD -64$^{\circ}$4322 & 9.3 & A2 Eu Sr & 7.4   
& 519.725 &4& $-$912 $\pm$ 142 \\
\end{tabular}
\end{center}
\end{table*}

In the following we give a brief overview of the previous knowledge of magnetic fields in the 
studied roAp 
stars. The most recent compilation of the measurements of magnetic fields in roAp stars is given 
by Mathys (\cite{M03}). Although the presence of a magnetic field is required by all the 
pulsation-driving mechanisms that have been proposed, knowledge of the magnetic fields of roAp 
stars is still very incomplete. Apart from HD\,6532, all stars in our sample are poorly studied. 
For six of the stars presented here the spectropolarimetric magnetic field 
measurements are the first to have been made. We note here that the previous magnetic field data 
mentioned below are presented in greater detail in Mathys \& Hubrig (\cite{MH97}) and Mathys 
\& Hubrig (2003, in preparation).

\subsection{Notes on individual stars}

{\it \object{HD\,6532}:}
The 6.9-min pulsations in this star were discovered by Kurtz \& Kreidl (\cite{KK85}). This star 
is singly periodic with a rotationally modulated frequency quintuplet (Kurtz et al. \cite 
{K96}). It is also the only star in our sample for which a short rotation period 
(P$_{\rm rot}=1.\!\!{^d}94497$) has been determined (see Kurtz et al. \cite{K96} and 
references therein). The spectral lines show significant Doppler broadening which complicates 
the magnetic field determination using high--resolution spectropolarimetry 
(Mathys \& Hubrig \cite{MH97}). An estimate of $v\,\sin i<47$~km~s$^{-1}$ has been determined 
from CASPEC spectra obtained with the ESO 3.6\,m spectrograph (Mathys \& Hubrig 2003, in 
preparation). The major advantage of using low-resolution spectropolarimetry with FORS\,1 is 
that polarization can be detected in relatively fast rotators as we measure the field in the 
hydrogen Balmer lines. 
Our determination of the longitudinal field yields a null value for this star.

{\it \object{HD\,9289}:}
The 10.5-min pulsations in this star were discovered by (Kurtz \cite{K93}). A later study 
(Kurtz et al. \cite{KMT94}) found the star to be multi-periodic with alternating even and odd 
$\ell$-modes, similar to the well-studied roAp star HR\,1217 (Kurtz et al. \cite{KWET02}). 
This star was observed only once with CASPEC 
(Mathys \& Hubrig 2003, in preparation), giving a nonzero longitudinal field of about 400~G. 
A longitudinal magnetic field of $(654\pm130)$~G has been diagnosed from the FORS\,1 spectra.

{\it \object{HD\,12932}:}
The 11.6-min pulsations in this star were discovered by Schneider \& Weiss (\cite{SW90}). 
This star is poorly studied, but appears to be singly periodic. A previous measurement with 
CASPEC (Mathys \& Hubrig 2003, in preparation) 
revealed a longitudinal field of 400~G. From the FORS\,1 spectra we derive 
$\left<{\cal B}_z\right>$ = $(638\pm169)$~G.

{\it \object{HD\,99563}:}
The 11.2-min pulsations in this star were discovered by Dorokhova \& Dorokhov (\cite{DD98}). 
The longitudinal magnetic field in this star at 2.1$\sigma$ level is reported here for the 
first time. 

{\it \object{HD\,122970}:}
The 11.1-min pulsations in this star were discovered by Handler \& Paunzen (\cite{HP99}) and 
studied in greater detail in two multi-site campaigns by Handler et al. (\cite{Han02}) who found 
two modes of different degree, $\ell$, in both data sets, along with frequency and amplitude 
modulation. 
A marginal hint of a longitudinal magnetic field in this star is reported here for the 
first time. Ryabchikova et al. (\cite {RS00}) determined a surface magnetic field of 2.3\,kG 
from the 
modelling of the rear--earth line profiles.

{\it \object{HD\.161459}:}
The 12.0-min pulsations in this star were discovered by Martinez \& Kauffmann (\cite{MK91}) and 
studied further by Martinez et al. (\cite{MKK91}). The presence of a magnetic field in this star 
is reported here for the first time. 

{\it \object{HD\,185256}:}
The 10.2-min pulsations in this star were discovered by Kurtz \& Martinez (\cite{KM95}). 
The presence of a magnetic field in this star is reported here for the first time. 

{\it \object{HD\,190290}:}
The multi-periodic 7.3-min pulsations in this star were discovered by Martinez \& Kurtz 
(\cite{MK90a}) and studied further by Martinez et al. (\cite{MKK91}). 
The presence of a magnetic field in this star is established here for the
first time, via two measurements.
This star shows the strongest longitudinal magnetic field (in absolute value) measured so far in 
any of the 32 known roAp stars. The two magnetic field measurements were separated by 4 days, 
but no significant variation was found between them.
 
{\it \object{HD\,193756}:}
The 13.0-min pulsations in this star were discovered by Martinez \& Kurtz (\cite{MK90b}). An 
attempt to detect the magnetic field in this star was made by Mathys \& Hubrig (1997), but no 
field was detected. We are not able to detect a longitudinal magnetic field on the 
FORS\,1 spectra.

{\it \object{HD\,196470}:}
The 10.8-min pulsations in this star were discovered by Martinez et al. (\cite{MKKJ90a}). 
The presence of a magnetic field in this star is reported here for the first time. 

{\it \object{HD\,203932}:}
The 5.9-min pulsations in this star were discovered by Kurtz (\cite{K84}). This is one of the 
shortest periods known for the roAp stars; only HD\,134214 has a shorter period at 5.7 min. 
Further study by Martinez et al. (\cite{MKH90b}) indicated the presence of some transient 
frequencies. Two previous attempts to measure the magnetic field in this star 
(Mathys \& Hubrig \cite{MH97}) yielded no detection. A longitudinal field of about $-150$~G was 
detected from a CASPEC spectrum by Mathys \& Hubrig (2003, in preparation). 
As the accuracy of our determination on FORS\,1 spectra is significantly worse than 
the accuracy achieved 
with CASPEC spectra for this star ($\approx$40\,Gauss), the longitudinal field could not be 
diagnosed.

{\it \object{HD\,213637}:}
The 11.5-min pulsations in this star were discovered by Martinez et al. (\cite{MMRE98}). 
Resolved magnetically split lines have been observed in this star by Mathys (\cite{M03}) and 
Kochukhov (\cite {KO03}).

{\it \object{HD\,217522}:}
The 13.7-min pulsations in this star were discovered by Kurtz (\cite{K83}). Further observations 
by Kreidl et al. (\cite{KKB91}) found another pulsation mode with a period of 8.3 min that was 
not present in the discovery data set, indicating transient modes, or strong amplitude 
modulation. The first magnetic observation of this star did not show any significant 
longitudinal magnetic field (Mathys \& Hubrig \cite{MH97}). Here we report the detection of a 
longitudinal field from the FORS\,1 spectra at the 4.1$\sigma$ level. 

{\it \object{HD\,218495}:}
The 7.4-min pulsations in this star were discovered by Martinez \& Kurtz (\cite{MK90b}). Both 
previous determinations of the longitudinal field by Mathys \& Hubrig (\cite{MH97}) yielded 
nonzero values at a low level of significance. A longitudinal field at the 6.4$\sigma$ level 
has been diagnosed from the FORS\,1 spectra.

\section{Discussion}\label{sec4}

The presence of non-zero mean longitudinal fields in 11 roAp stars gives an indication that 
magnetic fields of a 
few hundred gauss up to a few kilogauss are widespread among roAp stars. 
There are only three stars, HD\,6532, HD\,99563 and HD\,193756, for which a magnetic
field has not been detected at a level of $3\,\sigma$ or higher in the present and preceding 
studies.
Our results seem to hint that stronger magnetic 
fields tend to be found in hotter stars. Indeed, the stars of spectral type A0 and 
A2, HD\,190290, HD\,161459,
HD\,196470 and HD\,218495 show the largest longitudinal fields. 
Hubrig et al.\ (\cite{HNM00}) have 
already reported about the existence of such a trend in their study of the 
evolutionary state of 
magnetic Ap stars.
We should note that for roAp stars the effective temperatures 
derived from photometry are not in good agreement with the spectral classification 
(e.g., Kurtz \cite{K02}; North 2003, private communication). Because of 
the extremely anomalous 
energy distribution of 
these stars, the calibration of the photometric temperature indicators are
frequently questioned. The inconsistencies can only be resolved by adopting 
effective temperatures inferred from detailed spectroscopic studies.
In addition, because of the strong dependence of the longitudinal field on the rotational 
aspect, its usefulness to characterise actual field strength distributions is limited, but 
this can be overcome, at least in part, by repeated observations to sample various rotation 
phases, hence various aspects of the field. 

Two related outstanding problems in our understanding of the nature of the roAp stars are the 
pulsation driving mechanism and the mode selection mechanism (see Shibahashi (\cite{S03}) for 
discussion and references). The magnetic roAp stars pulsate only in high overtone modes while 
the non-magnetic $\delta$ Sct stars pulsate mostly in low overtone modes. Since these stars 
overlap in the H-R diagram, it is strongly suspected that the magnetic field plays a major 
role, whether directly or indirectly, in the driving and mode selection mechanisms. Except for 
HD\,122970 and HD\,217522, the stars in our sample do not have accurately determined Hipparcos 
parallaxes, hence their positions in the H-R diagram cannot be reliably determined. Nevertheless, in our recent study (Hubrig et al. \cite{HKMN00}) we were able to show that the domains of the 
roAp stars and the non-oscillating Ap (noAp) stars largely overlap in the H-R diagram. As this 
means that mass and internal structure differences between the roAp and noAp stars cannot be 
the only decisive factor in their respective evolution, the question arises whether another key 
factor plays a role in the excitation and mode selection mechanisms. The magnetic field is a 
likely candidate to be this factor. It is clearly necessary to understand the character of the 
surface magnetic fields in as much detail as possible.

These are the reasons why we are studying the magnetic field structure of the roAp stars. As 
mentioned by Mathys (\cite{M03}) there is an unusually high frequency of occurrence of strong 
magnetic fields and of slow rotation among roAp stars. Seven roAp stars with magnetically 
resolved lines have already been detected since the discovery of pulsations more than 20 years 
ago. However, quite a number of roAp stars have rather weak or undetectable longitudinal fields 
and strong quadratic fields (Mathys \& Hubrig \cite{MH97}). The determination of the quadratic 
field is based on the differential broadening that the field induces in unpolarized spectral 
lines having different magnetic sensitivities (for more details on the derivation of the 
quadratic magnetic field, see Mathys \cite{M99}). 
The rotational phase coverage achieved by the existing magnetic field data is insufficient and 
the structure of the magnetic fields in these roAp stars is presently unknown. Yet, from our 
results, it must be sufficiently tangled that it does not produce strong observable polarization 
signature. Clearly, further systematic studies of magnetic fields in these stars should be 
conducted with a view to derive the geometrical structure of the fields. To understand better 
the origin of pulsations the derived models of magnetic fields must then to be confronted with 
the pulsation properties of the roAp stars.

\begin{acknowledgements}
We are very grateful to S.\ Bagnulo for helpful discussions on data 
reduction. We would also like to thank P.\ North for the examination of the 
effective temperatures of the stars in our sample and to the anonymous referee for the 
constructive comments which helped us to improve the paper.
\end{acknowledgements}

\end{document}